\documentclass[11pt]{article}                               

\usepackage{latexsym}
\usepackage{amssymb}
\usepackage{amsmath}
\input{epsf}
\def\AFOUR{%
\setlength{\textheight}{9.0in}%
\setlength{\textwidth}{5.75in}%
\setlength{\topmargin}{-0.375in}%
\hoffset=-.5in%
\renewcommand{\baselinestretch}{1.17}%
\setlength{\parskip}{6pt plus 2pt}%
}

\AFOUR                                           %%%   ***

\makeatletter
\def\section{\@startsection {section}{1}{\z@}{-3.5ex plus -1ex minus
 -.2ex}{2.3ex plus .2ex}{\large\sc}}
\def\subsection{\@startsection{subsection}{2}{\z@}{-3.25ex plus -1ex minus
 -.2ex}{1.5ex plus .2ex}{\normalsize\sc}}
\makeatother

\makeatletter
\@addtoreset{equation}{section}

\makeatother

\newcommand{\nc}{\newcommand}
\newcommand{\rnc}{\renewcommand}

\nc{\bea}{\begin{eqnarray}}
\nc{\eea}{\end{eqnarray}}
\nc{\be}{\bea}
\nc{\ee}{\eea}

\rnc{\a}{\alpha}
\nc{\ab}{\bar{\a}}
\nc{\ap}{\a^{+}}
\nc{\abm}{\ab^{-}}
\rnc{\b}{\beta}
\nc{\bb}{\bar{\b}}
\nc{\bbp}{\bb_{\zb}^{+}}
\nc{\bm}{\b_{z}^{-}}
\nc{\oa}{\overline{\a}}
\nc{\ob}{\overline{\b}}
\rnc{\gg}{\gamma}
\rnc{\d}{\delta}
\nc{\f}{\phi}
\nc{\fb}{\bar{\phi}}
\nc{\vf}{\varphi}
\nc{\p}{\psi}

\rnc{\c}{\chi}
\nc{\la}{\lambda}
\nc{\m}{\mu}
\nc{\n}{\nu}
\rnc{\o}{\omega}
\nc{\Om}{\Omega}
\rnc{\t}{\theta}
\nc{\eps}{\epsilon}
\rnc{\S}{\Sigma}
\nc{\F}{\Phi}
\nc{\ldb}{\left(\!\left(}
\nc{\rdb}{\right)\!\right)}

\nc{\trac}[2]{{\textstyle\frac{#1}{#2}}}

\nc{\ex}[1]{\mbox{e}^{\,\textstyle#1}}

\nc{\mat}[4]{\left(\begin{array}{cc}#1&#2\\#3&#4\end{array}\right)}

\nc{\som}[9]{\left(\begin{array}{ccc}#1&#2&#3\\#4&#5&#6\\#7&#8&#9%
\end{array}\right)}

\nc{\tr}{\mathop{\mbox{tr}}\nolimits}
\nc{\ad}{\mathop{\mbox{ad}}\nolimits}
\nc{\Tr}{\mathop{\mbox{Tr}}\nolimits}
\nc{\Det}{\mathop{\mbox{Det}}\nolimits}
\nc{\rk}{\mathop{\mbox{rk}}\nolimits}
\nc{\ra}{\rightarrow}
\nc{\Ra}{\Rightarrow}
\nc{\LRa}{\Leftrightarrow}
\nc{\ot}{\otimes}
\rnc{\ss}{\subset}
\nc{\nul}{\noindent\underline}
\nc{\non}{\nonumber\\}
\nc{\ZZ}{\mathbb{Z}}
\nc{\RR}{\mathbb{R}}
\nc{\CC}{\mathbb{C}}

\nc{\subs}[1]{{\vspace*{0.5cm}}%
{\noindent\underline{#1}}%
{\vspace*{0.3cm}}}
\nc{\zb}{\bar{z}}
\rnc{\lg}{\mathfrak{g}}
\nc{\lt}{\mathfrak{t}}
\nc{\lk}{\mathfrak{k}}
\nc{\lh}{\mathfrak{h}}
\nc{\pik}{\Pi_{\lk}}
\nc{\pip}{\Pi_{+}}
\nc{\pim}{\Pi_{-}}
\nc{\pih}{\Pi_{\lh}}
\nc{\jz}{J_{z}}
\nc{\jzh}{\jz^{\lh}}
\nc{\jzp}{\jz^{+}}
\nc{\jzm}{\jz^{-}}
\nc{\del}{\partial}
\nc{\dz}{\del_{z}}
\nc{\dzb}{\del_{\bar{z}}}
\nc{\az}{A_{z}}
\nc{\azb}{A_{\bar{z}}}
\nc{\g}{g^{-1}}
\nc{\dw}{\Delta_{W}}
\nc{\Ad}{{\mbox{Ad}}}
\nc{\ks}{Ka\-za\-ma-\-Su\-zu\-ki}
\nc{\KS}{\ks}
\nc{\ksm}{\ks\ model}
\rnc{\AA}{{\Bbb A}}
\nc{\BB}{{\Bbb B}}
\nc{\PP}{{\Bbb P}}
\nc{\cpm}{\CC\PP(m)}
\nc{\cpn}{\CC\PP(n)}
\nc{\cp}[1]{\CC\PP(#1)}
\nc{\gmn}{G(m,m+n)}
\nc{\gmnk}{\gmn_{k}}
\nc{\cO}{{\cal O}}
\nc{\bcO}{\bar{\cO}}
\nc{\bO}{\bar{O}}
\nc{\oQ}{\overline{Q}}

\begin{document}
\global\parskip=4pt

%%%%%%%%% title page %%%%%%%%%%%%%%%%%%%%%%%%%%%%%%%%%%%%%%%%
\begin{titlepage}
\begin{center}
\vspace*{.5in}
{\LARGE\bf Chern-Simons Theory on Seifert 3-Manifolds}\\[.3in] 
%{\Large\bf Abelianisation and q-deformed Yang-Mills Theory}\\
\vskip 0.7in
{\bf Matthias Blau}\footnote{e-mail: blau(at)itp.unibe.ch}
\vskip .1in
Albert Einstein Center for Fundamental Physics, Institute of
Theoretical Physics, Bern University, Switzerland.
\vskip 0.2in
%and
\vskip 0.2in
{\bf George Thompson}\footnote{e-mail: thompson(at)ictp.it}
\vskip .1in
ICTP, 
 Trieste, 
Italy.\\

\end{center}
\vskip .4in
\begin{abstract} 
\noindent 
We study Chern-Simons theory on 3-manifolds $M$ that are circle-bundles
over 2-dimensional orbifolds $\Sigma$ by the method of
Abelianisation. This method, which
completely sidesteps the issue of having to integrate over the moduli
space of non-Abelian flat connections, reduces the complete partition
function of the 
non-Abelian theory on $M$ to a 2-dimensional Abelian theory on the
orbifold $\Sigma$, which is easily evaluated. 

\end{abstract}

\end{titlepage}

%%%%%%%%% end of title page %%%%%%%%%%%%%%%%%%%%%%%%%%%%%%

%\vspace*{1cm}
\begin{small}
%\tableofcontents
\end{small}

%\newpage
\setcounter{footnote}{0}

\section{Introduction}

Chern-Simons theory \cite{WCS} has been with us now for about 25
years. The Chern-Simons path integral, at every level $k \in
\mathbb{Z}$ and for Lie group $G$,
\be
Z_{CS}[M,G]= \int_{\mathcal{A}}\,
\exp{\left(i\frac{k}{4\pi}\int_{M}\Tr{AdA + \frac{2}{3}A^{3} }\right)} 
\ee
gives us a (framed) invariant of the 3-manifold $M$. Witten \cite{WCS}
and Reshetikhin and Turaev \cite{RT} gave
surgery prescriptions for these invariants (the first based on
conformal field theory, the second on quantum groups). 

Very early on
Freed and Gompf \cite{FG} expressed the invariant for Seifert
3-manifolds and the group $G=SU(2)$ in terms of the $S$ and $T$
matrices of conformal field theory. Jeffrey \cite{Jeffrey} obtained
rather more explicit formulae for Lens spaces. Lawrence and
Rozansky \cite{RL} obtained just as explicit results for Seifert
rational homology spheres ($\mathbb{Q}$HS's). Mari\~{n}o
\cite{marino} extended the results of \cite{RL} to compact simply-laced
$G$. Interestingly enough Lawrence and
Rozansky and Mari\~{n}o were predominantly interested in obtaining
asymptotic formulae around the (isolated) trivial connection from the
exact result.

Unlike the surgery prescription, strategies for
the exact
evaluation of the path integral formulation of Chern-Simons theory
are few and far between. There have been many studies of the
perturbative aspects of the theory from the path integral view point,
unfortunately
far too many to review here. But, as already mentioned, there is
a dearth of exact evaluations based directly on the path
integral. There are some exceptions to this however. One such
exception is due to
Jeffrey \cite{Jeffrey} who evaluated the partition function of
Chern-Simons in the semi-classical approximation for mapping
tori (at least formally). Another is our evaluation of the path
integral on 3-manifolds of 
the form $\Sigma \times S^{1}$ for $\Sigma$ a genus $g$ Riemann
surface \cite{btg/g}. 

Somewhat more recently Beasley and Witten
\cite{BW} have developed a localisation procedure for the path
integral that extracts the contribution around isolated
connections. This method, based on non-Abelian localisation
\cite{W2dRev}, requires a contact structure to be chosen on the
3-manifold and, for calculations, a $U(1)$ action is also
required. The two requirements essentially fix one to Seifert
$\mathbb{Q}$HS's. Beasley \cite{Beasley} has extended the approach to
include the expectation value of Wilson loops along the $U(1)$ fibre.

In \cite{BTcs} we were able to extend the diagonalisation techniques
introduced in \cite{btg/g} to manifolds which are circle
bundles over smooth Riemann surfaces, in particular to the Lens spaces
$L(p,1)$. This work has some similarity to \cite{BW,Beasley} but
perhaps the biggest difference is that while these authors obtain
contributions about particular connections our technique evaluates the
complete path integral. Furthermore, the formulae obtained, unlike
those that come from a semi-classical approximation, do not involve
complicated integrals over moduli spaces of flat connections but
rather integrals over the Cartan 
subalgebra of the gauge group.

The present paper is a continuation of \cite{BTcs} to 3-manifolds which are
$U(1)$ bundles over orbifolds. One motivation for the present study
is, then, to apply the procedure of Abelianisation in the
case that the smooth 3-manifold is a circle V-bundle over an
orbifold. Understanding the correct condition when diagonalising in
this context is the main technical difficulty.

Apart from the intrinsic importance of being able to evaluate the path
integral for the Chern-Simons partition function, there is also the
benefit from the possibility to use the techniques in other situations. In
particular if there is no obvious, or perhaps obviously practical,
surgery prescription, then other means are needed to glean non-perturbative
information. There are a number of such situations which we will
address elsewhere. These include:
\begin{itemize} 
\item Three dimensional $BF$ theory which is an example of a theory
for which there is no known surgery prescription and to which our
methods apply. Such theories are of interest because of their
relation to gravity \cite{WBF}.
\item The $N_{T}=2$
topological supersymmetric extension of $BF$ theory on Seifert
$\mathbb{Q}$HS's. These theories are presentations of the Casson invariant
and its generalisations \cite{BTCasson}.
\item Yang-Mills theories on 2-dimensional orbifolds. Even though these are
  related to Yang-Mills theories on smooth surfaces with `parabolic
  points' \cite{btictp} they may be of independent interest. 
\end{itemize}

Another motivation comes from the fact that the 
Chern-Simons partition function on particular Seifert manifolds is the
same as certain 
intersection pairings on spaces of connections on a Riemann
surface. This relationship then gives a geometric meaning to the
Chern-Simons invariants of these manifolds (and of knots in them).
Certainly one application is a slightly different geometric
understanding of the partition function as intersection pairings on
the infinite dimensional space of connections modulo the gauge
group. This can be established by making use of the topological supersymmetric
extension of Chern-Simons introduced in \cite{thompson}. 

There is another use for the toplogical supersymmetric extension of
\cite{thompson}. Namely, K\"{a}ll\'{e}n \cite{Kallen} views the
Chern-Simons action as an observable within the topologically twisted
Yang-Mills theory and then uses cohomological localisation to
reproduce \cite{BW} around the trivial connection. Ohta and Yoshida
\cite{OY} combine the
topologically twisted Yang-Mills theory of \cite{Kallen} and
diagonalisation \cite{btg/g, BTcs} to evaluate the path integral for
various supersymmetric Yang-Mills Chern-Simons theories.

The contents of this paper are as follows:

We begin with the formula, due to  Lawrence and Rozansky \cite{RL},
for the partition function of $SU(2)$ Chern-Simons theory on a Seifert
$\mathbb{Q}$HS and its generalisation due to Mari\~{n}o
\cite{marino}. We show how these formulae can be written in various suggestive
ways as integrals over the Cartan subalgebra of the group in
question. Such a formulation is a preparation for formulae that arise
on evaluating the path integral by diagonalisation.

Then we turn to a brief description of Seifert 3-manifolds as $S^{1}$
V-bundles over 2-dimensional orbifolds. In particular we introduce the
conditions that a Seifert manifold be a $\mathbb{Q}[g]$HS which has,
apart from an extra $\mathbb{Z}^{2g}$ summand in $\mathrm{H}_{1}(M)$,
the homology of a $\mathbb{Q}$HS. We are able to evalute the
Chern-Simons path integral on this class of Seifert manifolds.

Next we come to the crux of the matter, namely diagonalising a
component of the gauge connection so that it lies in the Cartan
subalgebra of the Lie algebra of the gauge group. On doing this we
are left with a sequence of Gaussian integrals to perform and a price
to be paid. That price is that the abelianised field is correctly
thought of as a section of certain line bundles over the orbifold and
part of our task is to determine which line bundles.

Once the bundles that arise on
Abelianisation are clear, apart from some arithmetic the 
evaluation of the Chern-Simons path integral is almost identical to
that presented in \cite{BTcs} and so we will be rather brief about the
details. As noted above, we are basically evaluating some Gaussian path
integrals which give rise to determinants. The evaluation proceeds in
a sequence of steps. Firstly we split the functional determinants into
their absolute value and the phase. Then we give a zeta function and
eta function type regularisation of these. An application of the
Riemann-Roch-Kawasaki index theorem \cite{Kawasaki} (the extension of the Riemann-Roch theorem
to V-manifolds or orbifolds) allows us to push the
calculation down to the orbifold. Finally, we can make use of an
orbifold version of Hodge theory \cite{Baily} to evaluate the resulting Abelian
theory. 

The last section deals with the evaluation of the expectation value of
particular Wilson loops. These are the lines that wrap around the
$S^{1}$ fibration of $M$. This is quite straightforward to do as these
Wilson loops do not interfere with the previous method of evaluation
as the Gaussian nature of the model is maintained.

There are two appendices. The first gives an example of
diagonalisation of a smooth section of a (smooth) $SU(2)$ bundle over
$M$ and what that means for summing bundles over the orbifold base of
$M$. This is intended to motivate the choices made in the body of the
paper. In the second appendix we give the generators and relations for
$\pi_{1}(M)$ and explicit forms for their abelianisation. Along the
way we also give an example of an irreducible non-Abelian connection
of the type that we do not have to take into account in our evaluation
of the path integral.

\section{The Formulae of Lawrence-Rozansky and Mari\~{n}o}\label{sec2}

The formula found by Lawrence and Rozansky \cite{RL} for the partition
function of Chern-Simons theory of a Seifert
$\mathbb{Q}$HS $M$, up to an overall constant, is
\be
Z_{CS}(M,\, SU(2))=\sum_{r = -Pk_{\lg}}^{Pk_{\lg}} \ex{-\frac{i
    \pi d}{2Pk_{\lg}}r^{2}} \left( \ex{\frac{i
    \pi r}{k_{\lg}}}- \ex{-\frac{i
    \pi r}{k_{\lg}}}\right)^{2-N}\, \prod_{i=1}^{N} \left(\ex{\frac{i
    \pi r}{a_{i}k_{\lg}}}- \ex{-\frac{i
    \pi r}{a_{i}k_{\lg}}} \right) \label{LR1}
\ee
where the $a_{i}$ for $i=1, \dots , N$ are part of the data of a Seifert manifold $M$ 
(see section \ref{secseif} for more details) and
$P=\prod_{i=1}^{N}a_{i}$, while $|d|$ is the order of
$\mathrm{H}_{1}(M,\mathbb{Q})$ (with $d= \pm |d|$ corresponding to the two choices 
of orientation). We generally denote
$k_{\lg}=k+c_{\lg}$ where $c_{\lg}$ is the dual 
Coxetor number for the group $G$. In this formula, and its
generalisation to other simply-laced groups $G$ (see \cite{marino} and
the discussion just after (4.9) there), there
are restrictions on the summation. Here we see that for $N>2$
\be
\left( \ex{\frac{i
    \pi r}{k_{\lg}}}- \ex{-\frac{i
    \pi r}{k_{\lg}}}\right)^{2-N}\nonumber
\ee
diverges whenever $k_{\lg} | r$ and it is these points which are
discarded in the sum. This is the analogue of a similar issue that
arose in the path integral derivation of the Verlinde formula for the
dimension of the space of conformal blocks \cite{btg/g} and we deal
with it in the same way here, as we describe below.

As in \cite{BTcs} we introduce a gauge invariant partition function
\be
Z_{q,P,d}(f) = \frac{q^{\mathbf{rk}}}{|W|\, V}\, \sum_{s\in
  \mathbb{Z}^{\mathbf{rk}}}  \int_{\lt} f(\phi) \,
\exp{\left(i\frac{q}{4\pi }\frac{d}{P} \Tr{\phi^{2}} + iq \Tr{s
      \phi}\right)} \label{parent}
\ee
Here $W$ is the Weyl group which acts by permutation on the Cartan
elements, $\mathbf{rk}$ is the rank of the group $G$, 
$q$ is an integer (later to be identified with $k_{\lg}$), 
$f(\phi)$ is any
function which is invariant under both the shift 
$\phi \rightarrow \phi + 2\pi P$ and the action of $W$ and
$V=\mathbf{Vol}(d\mathbb{Z}^{\mathbf{rk}}) $. We have set
$\phi = \sum_{i} \phi^{i}\alpha_{i}$ and $s=\sum_{i} s^{i}\alpha_{i}$
where $\alpha_{i}$ are simple roots of a group $G$ (for simplicity in
the following we consider $G$ to be simply-laced). The gauge symmetry that
is enjoyed by this partition function is, with
$n=\sum_{i}n^{i}\alpha_{i}$ and the $n^{i} \in \mathbb{Z}$,
\be
\phi \rightarrow \phi + 2\pi n P, \;\;\; s \rightarrow s - d n \label{shifts}
\ee
Note that this says that we may shift $\phi/P$
by elements of the integral lattice $I$. The discrete group that acts is
then the affine Weyl group $\Gamma^{W}= I \rtimes W$.

Now, using $I$ we can either gauge fix $\phi$ to lie
between $-\pi P\leq  \phi \leq \pi P$ or we can gauge fix $s$ so that $s
\in \mathbb{Z}_{d}$, or we can use the whole affine Weyl group to
restrict $\phi$ to $\lt/\Gamma^{W}$. Therefore  we arrive at the equalities
\bea
Z_{q,P,d}(f) & = & \frac{1}{|W|}\sum_{s \in \mathbb{Z}^{\mathbf{rk}}} \int_{-\pi P
  q}^{\pi Pq}\dots \int_{-\pi P
  q}^{\pi Pq} f(\phi/q) \, 
\exp{\left(\frac{i}{4\pi}\frac{d}{Pq} \Tr{\phi^{2}} + i\Tr{s \phi}
    \right)} \nonumber\\
& = & \frac{1}{|W|}\sum_{s\in \mathbb{Z}_{d}}\int_{\lt} f(\phi/q)
\,\exp{\left(\frac{i}{4\pi}\frac{d}{Pq} \Tr{\phi^{2}} + i\Tr{s \phi} 
    \right)} \nonumber\\
& = & \sum_{s\in
  \mathbb{Z}^{\mathbf{rk}}}  \int_{\lt/\Gamma^{W}} f(\phi) \,
\exp{\left(i\frac{q}{4\pi }\frac{d}{P} \Tr{\phi^{2}}  + iq \Tr{s
      \phi}\right)} \label{z}
\eea
The sum over $s$ in the first of the equalities of (\ref{z}) fixes
$\phi = r\pi$ for $r \in \mathbb{Z}^{\mathbf{rk}}$,  while the range of integration
restricts each of the possible integers $r$ to lie in $-Pq \leq r \leq Pq$.
Consequently we have
\be
Z_{q,P,d}(f) = \frac{1}{|W|}\sum_{r \in \mathbb{Z}^{\mathbf{rk}}/2Pq
  \mathbb{Z}^{\mathbf{rk}} } f(\pi r/q) \,
\exp{\left(i\frac{\pi}{4}\frac{d}{Pq}\Tr{r^{2}}\right) }
\ee
A suitable choice of the function $f$ then reproduces the formula
(\ref{LR1}) on taking $q=k_{\lg}= k+c_{\lg}$. More generally, we let
$f = \sqrt{T_{M}(\phi, \, a_{1}, \dots , a_{N})}$ (the positive root) where
\be
 T_{M}(\phi;\, a_{1}, \dots , a_{N}) =
T_{S^{1}}(\phi)^{2-2g-N}
. \prod_{i=1}^{N}T_{S^{1}}(\phi/a_{i}) \label{M-Ray-Singer}
\ee
to reproduce the formulae of Mari\~{n}o \cite{marino}. This formula
relates the Ray-Singer torsion of $M$, $T_{M}(\phi, \, a_{1}, \dots ,
a_{N})$ to $T_{S^{1}}(\phi/a)$ which is the 
Ray-Singer torsion on the circle (modulo $\mathbb{Z}_{a}$) evaluated
at a flat connection $(\phi/a)\, d\theta$. 

Our evaluation of the Chern-Simons path integral will eventually 
lead us to an expression of the form \eqref{parent}, with $f(\phi)$
precisely as above, and at that
point we can appeal to the above discussion to establish the 
connection with the results of \cite{RL,marino}.

\section{Seifert 3-Manifolds}
\label{secseif}

We will consider Chern-Simons theory on 3-manifolds $M$ which are
themselves principal $U(1)$ V-bundles $U(1) \rightarrow M
\stackrel{\pi}{\rightarrow} \Sigma$ over 2-dimensional orbifolds
$\Sigma$ of genus $g$. We suppose that there are $N$ orbifold points
on $\Sigma$ which are modeled on
$\mathbb{C}/\mathbb{Z}_{a_{i}}$. A line V-bundle, away from orbifold
points is characterised as an ordinary line bundle but is given, in the
neighbourhood of the $i$-th orbifold point, by the identification on $D
\times \mathbb{C}$,
\be
(z, w) \simeq (\zeta.z,\, \zeta^{b_{i}}.w),\;\;\; \zeta = \exp{(2\pi
  i/a_{i}) } \nonumber
\ee 
where $D \subset \mathbb{C}$ is a disc centred at the orbifold point.

$M=S(\mathcal{L})$ is then a circle V-bundle
associated to a line V-bundle $\mathcal{L}$. The data which
goes into specifying $M$ is $[\deg{(\mathcal{L})},\, g,\,
(a_{1},b_{1}) , \dots ,(a_{N},b_{N})] $. 

The Seifert manifold $M$ is smooth if
\be
\mathrm{gcd}(a_{i},b_{i}) = 1, \;\; \mathrm{for}\; i=1, \dots
, N \nonumber
\ee
which means, in particular, that the $b_{i} \neq 0$. The Seifert
manifold will be an $\mathbb{Z}$HS (integral homology 
sphere) iff the line bundle
$\mathcal{L}_{0}$ that defines it satisfies
\be
g=0, \;\; c_{1}(\mathcal{L}_{0}) = \pm \prod_{i=1}^{N}
\frac{1}{a_{i}} \label{int-hom-sph}
\ee
The last condition implies that the numbers $a_{i}$ be pairwise
relatively prime\footnote{From (\ref{chern-notation}) and
  (\ref{int-hom-sph}) we have that for 
$M$ to be a $\mathbb{Z}$HS that $(\prod_{i}a_{i}) . (n + \sum_{j}
b_{j}/a_{j}) = \pm 1$ where $n$ is the degree of the bundle that defines
$M$. Now suppose that the greatest common divisor of two of the $a_{i}$
is not unity and, by re-ordering if required, let those two be $a_{1}$
and $a_{2}$ and such that $a_{2}= ta_{1}$ ($t \in
\mathbb{Z}_{>0}$). The equation to be solved becomes $a_{1}. m =\pm 1$
with $m= (\prod_{j\geq 3}a_{j})[t a_{1} (n + \sum_{i\geq 3}
b_{i}/a_{i}) + tb_{1}+ b_{2}]$ and clearly $m \in \mathbb{Z}$ so there
is no solution. Consequently, for $M$ to be a $\mathbb{Z}$HS the
$a_{i}$ must be pairwise relatively prime.}  so that one has the arithmetic condition
\be
\mathrm{gcd}(a_{i},a_{j}) = 1, \;\; i\neq j \nonumber
\ee

As an example the Poincar\'{e} $\mathbb{Z}$HS $M=\Sigma(2,3,5)$
has
the two possible descriptions $[-1,0,(2,1),(3,1),(5,1)]$ with
$c_{1}(\mathcal{L}_{0})=1/(2.3.5)$ and $[-2,0,(2,1),(3,2),(5,4)]$ with
$c_{1}(\mathcal{L}_{0})=-1/(2.3.5)$. Quite generally, if
$[\deg{(\mathcal{L})},\, g,\,(a_{1},b_{1}) , \dots ,(a_{N},b_{N})]$ is
a manifold with $c_{1}(\mathcal{L}_{0})=\pm 1/(a_{1}\dots a_{N})$,
then $[-\deg{(\mathcal{L})}-N,\, g,\,(a_{1},a_{1}-b_{1}) , \dots
,(a_{N},a_{N}-b_{N})]$ is one with $c_{1}(\mathcal{L}_{0})=\mp
1/(a_{1}\dots a_{N})$ (we are taking the inverse line
bundle)\footnote{That the degree in the examples is always negative is
  not an accident. The equation to solve is $n (\prod_{j}a_{j}) + s =
\pm 1 $ with $s = (\prod_{j}a_{j}).\sum_{i}b_{i}/a_{i}$ a positive
integer so that $n\leq 0$ given that the $a_{i}\geq 2$ and the
$b_{i}\geq 1$.}.

If one takes $M$ to be the total space of the circle bundle of
$\mathcal{L}_{0}^{\otimes d}$, rather than that of $\mathcal{L}_{0}$,
then $M$ is a $\mathbb{Q}$HS (rational homology sphere) with 
\be
|d|= |\mathrm{H}_{1}(M, \mathbb{Z})| \nonumber
\ee
In both of these cases, as the $a_{i}$ are mutually coprime, all
line V-bundles on $\Sigma$ are some tensor power of $\mathcal{L}_{0}$.

The Gysin sequence played an important role in our previous evaluation
of the path integral on $U(1)$ bundles over smooth curves allowing us
to count $U(1)$ bundles over the total space which are pullbacks from
the base. Likewise, we would like to know the image of the pullback map
\be
\mathrm{Pic}(\Sigma) \stackrel{\pi^{*}}{\longrightarrow}
[\mathrm{Line\;bundles\;over}\;M] \stackrel{c_{1}}{\longrightarrow}
\mathrm{H}^{2}(M, \mathbb{Z}) \nonumber
\ee
Fortunately there is a Gysin sequence for $U(1)$ V-bundles over 2-dimensional
orbifolds \cite{FS} which gives us the required information. The
required result is part of Theorem 2.3 in \cite{FS} (see also Remark 2.0.20
in \cite{MOY}) for $M$ smooth,
\be
\mathrm{H}^{2}(M, \mathbb{Z}) \cong \mathrm{Pic}(\Sigma)/\mathbb{Z}[\mathcal{L}]
\oplus \mathbb{Z}^{2g}\label{gysin}
\ee
where $M=S(\mathcal{L})$. 
It is the subgroup $\mathrm{Pic}(\Sigma)/\mathbb{Z}[\mathcal{L}] \subset
\mathrm{H}^{2}(M, \mathbb{Z})$ which is the image of the pullback map
and when $c_{1}(\mathcal{L}) \neq 0$ this is finite and Abelian. When
$M$ is a $\mathbb{Q}$HS, then
$\mathcal{L}=\mathcal{L}_{0}^{\otimes d}$ and $\mathrm{Pic}(\Sigma)=
\mathbb{Z}[\mathcal{L}_{0}]$ so that
$\mathrm{Pic}(\Sigma)/\mathbb{Z}[\mathcal{L} ] \cong
\mathbb{Z}_{d}$.

\section*{Choice of Manifold}

The technique that we make use of is Abelianisation. In particular, we
diagonalise sections $\phi : M \longrightarrow \ad{G}$ of the adjoint
bundle which are constant along the fibres of $M$, 
in the sense of conjugating them into maps into the Cartan sublagebra.
For reasons 
disucssed below we consider the case that $G$ is simply-connected, 
so that $\ad{G}\cong M \times \lg$.

However, even in this case there are topological 
obstructions to Abelianisation and, if one insists on diagonalising anyway, the
price to be paid is the `liberation' of non-trivial line bundles on
the base of the fibration $S^{1} \longrightarrow M\longrightarrow
\Sigma$. An important part of the technique, then,
is to be able to determine which line bundles we will need to
count. In the case of trivial bundles $M
= S^{1} \times \Sigma$ over a smooth Riemann surface 
we find that we must count all possible line bundles on
$\Sigma$. In the case of 
nontrivial bundles, the circle bundle of a non-trivial line bundle
$\mathcal{L}$, 
over a smooth Riemann surface one counts the $c_{1}(\mathcal{L})$ available
torsion bundles (these arise as $\pi^{*}(\mathcal{L}^{\otimes c_{1}(\mathcal{L})})=
\mathcal{O}$). 

Hence, we need to be able to follow the line bundles which are
available. This is most easily done if there is only one generator
that pulls back to the 3-manifold. This is the case in the examples of
the previous paragraph. Other examples include particular smooth
Seifert manifolds which are constructed as follows. Let
$\Sigma$ be a genus $g$ Riemann surface with $N$ orbifold
points $\{p_{i}\}$ such that the isotropy data
$a_{i}$ at the points $p_{i}$ are relatively prime
$\gcd{(a_{i}, \, a_{j})}=1$ for $i\neq j$. As the line bundle
$\mathcal{L}_{0}$ with $c_{1}(\mathcal{L}_{0})= \prod (a_{i})^{-1}$ generates the
Picard group of orbifold line bundles on $\Sigma$
(\ref{gysin}) the pullback to
$S(\mathcal{L}_{0})$ of any orbifold line bundle is trivial. (This is
the orbifold 
analogue of the fact that the pullback of any line bundle on $S^{2}$ to $S^{3}$
is trivial.) However, there is an important caveat. The $G$ bundle
that we started with is a smooth bundle and can be thought of as the
pullback of an honest $G$ bundle on $\Sigma$, i.e.\ one with
trivial isotropy data at the orbifold points. Consequently the line 
bundles that appear on diagonalisation must be honest line bundles
(there is no special discrete action over the orbifold points). All
such line bundles on $\Sigma$ are powers of
$\mathcal{L}_{P}=\mathcal{L}_{0}^{\otimes P}$ where 
\be 
P= \prod_{i=1}^{N}a_{i} \label{P}
\ee
and it is these line bundles, if we are to sum, that
we should sum over (though they all pullback to the trivial line bundle).

Notice, in the above discussion, that for diagonalisation we do not
need the extra 
condition that the genus of the Riemann surface vanish, so we are not
only dealing with $\mathbb{Z}$HS's (\ref{int-hom-sph}). For brevity we
will denote those $M= S(\mathcal{L}_{0})$ by $\mathbb{Z}[g]$HS's when
we relax the condition on the genus.

On the other hand the manifold $S(\mathcal{L}_{0}^{\otimes d})$ is
such that there are $d$ torsion bundles available and, on
diagonalising, we would have to count these (the $S(\mathcal{L}_{0}^{\otimes
  d})$ having the same relationship to $S(\mathcal{L}_{0})$ as the Lens spaces
have to $S^{3}$). As we have seen, in order to keep track of the fact
that our $G$ bundle is smooth we should consider the torsion bundles
to be of the form $\mathcal{L}_{P}^{\otimes m}$ for $m \in
\mathbb{Z}_{d}$. Once more we may  also consider that $g \neq 0$ and
we denote such manifolds as $\mathbb{Q}[g]$HS's.

\section*{Degree and First Chern Number}

There is some disparity in the literature regarding the
nomenclature used with regards to Chern classes, degree and so on. We
adopt the notation that 
\be
c_{1}(\mathcal{L}) = \deg{(\mathcal{L})} +\sum_{i=1}^{N}
\frac{b_{i}(\mathcal{L})}{a_{i}}
\label{chern-notation}
\ee
where the degree $\deg{(\mathcal{L})}$ is an integer and the isotropy weights
$b_{i}(\mathcal{L})$ each satisfy
\be
0 \leq b_{i}(\mathcal{L}) < a_{i} \nonumber
\ee
for every line bundle $\mathcal{L}$. Hence, with our definition,
$c_{1}(\mathcal{L}) \in \mathbb{Q}$.
Note that
$\deg(\mathcal{L})=c_{1}(|\mathcal{L}|) \in \mathbb{Z}$ where
$|\mathcal{L}|$ is the 
associated (smooth) line bundle on the smooth curve $|\Sigma|$ (by
smoothing the orbifold points and taking no isotropy there,
$b_{i}(\mathcal{L})=0$).

One way to think about this is as follows: A line bundle is
equivalent to a divisor, which in this case is a (smooth) point on
$\Sigma$. Each smooth point comes with weight one. A
degree $n$ line bundle is the same as the sum of $n$ divisors (say $n$
times one divisor). The orbifold points $\{ p_{i} \}$ have weight
$1/a_{i}$ and so they correspond to line V-bundles with `degree'
$1/a_{i}$. If one considers the divisor $a_{i}.\{p_{i}\}$ (which is
now like a smooth point) then the
associated line V-bundle is a line bundle and its degree feeds into $\deg$.

A fact which will be important for
us later is that while the first Chern class behaves well under tensor
product,
\be
c_{1}(\mathcal{L}\otimes \mathcal{K}) = c_{1}(\mathcal{L}) +
c_{1}(\mathcal{K}) \nonumber 
\ee
the degree does not. Rather, one has from this formula and the
definition,
\be
c_{1}(\mathcal{L}\otimes \mathcal{K}) = \deg{(\mathcal{L}\otimes \mathcal{K})} + \sum_{i=1}^{N}
\frac{b_{i}(\mathcal{L}\otimes \mathcal{K})}{a_{i}} \nonumber
\ee
with $0 \leq b_{i}(\mathcal{L}\otimes \mathcal{K} ) < a_{i}$ that
\be
\deg{(\mathcal{L}\otimes \mathcal{K})} = \deg{(\mathcal{L})} +
\deg{(\mathcal{K})} + \sum_{i=1}^{N} \left\lfloor 
  \frac{b_{i}(\mathcal{L})+ b_{i}(\mathcal{K})}{a_{i}} \right\rfloor \label{LxK}
\ee
where $\lfloor x \rfloor$ is the floor function
\be
\lfloor x \rfloor = \max \left\{ n\in \mathbb{Z}\,  |\,  x \geq n
\right\} \label{intpart} 
\ee
and is such that
\be
\lfloor -x \rfloor = \left\{ \begin{array}{ll}
-1 - \lfloor x \rfloor & x \in \mathbb{R}\backslash \mathbb{Z}\\
 - \lfloor x \rfloor & x \in \mathbb{Z}
\end{array} \right. \label{-intpart}
\ee
and where the isotropy weights satisfy
\be
b_{i}(\mathcal{L}\otimes \mathcal{K}) = \left( b_{i}(\mathcal{L})+
  b_{i}(\mathcal{K}) \right)
\mod  a_{i}, \;\; 0 \leq b_{i}(\mathcal{L}\otimes \mathcal{K}) < a_{i}\nonumber
\ee

Introduce the symbol $((.))$ defined by
\be
((x)) = \left\{ \begin{array}{ll}
x-\lfloor x\rfloor -\frac{1}{2}, & x \in \mathbb{R}\backslash \mathbb{Z} \\
0, & x \in \mathbb{Z} 
\end{array} \right. \nonumber
\ee
which has unit period $((x +1))=((x))$ and is odd under change of sign
$((-x))=-((x))$.

For any line bundle $\mathcal{L}$, as $0 \leq b_{i}(\mathcal{L}) < a_{i}$,
\be
b_{i}(\mathcal{L}^{-1}) = \left\{ \begin{array}{ll}
a_{i}- b_{i}(\mathcal{L}), & b_{i}(\mathcal{L}) \neq 0 \\
0, & b_{i}(\mathcal{L})=0
\end{array}
\right.
\ee
so that
\be
\frac{b_{i}(\mathcal{L})-b_{i}(\mathcal{L}^{-1})}{a_{i}} =
\left\{ \begin{array}{ll}
2b_{i}(\mathcal{L})/a_{i}-1 ,&  b_{i}(\mathcal{L}) \neq 0\\
0, & b_{i}(\mathcal{L})=0
\end{array}
\right. =2\, ((b_{i}(\mathcal{L})/a_{i}))\nonumber
\ee
since $\lfloor b_{i}(\mathcal{L})/a_{i} \rfloor =0$.
Consequently,
\be
\deg{(\mathcal{L})}-\deg{(\mathcal{L}^{-1})} = 2\, c_{1}(\mathcal{L}) - 2
\sum_{i=1}^{N}((b_{i}(\mathcal{L})/a_{i})) \nonumber
\ee
This trick we took from \cite{Nic}.

Likewise, providing that
$\gcd{(a_{i}, b_{i}(\mathcal{L}))}=1$,
\be
\deg{(\mathcal{L}^{\otimes n})} + \deg{(\mathcal{L}^{-\otimes n})} =  - N +
\sum_{i=1}^{N}\phi_{a_{i}}(n) \nonumber 
\ee
where 
\be
\phi_{a_{i}}(n)= \left\{ \begin{array}{ll}
1 & \mathrm{if}\; a_{i}| n\\
0 & \mathrm{otherwise} 
\end{array} \right. \label{phifunction}
\ee
is a function introduced in  \cite{BW}. Notice that the function
$\phi_{a_{i}}(n)$ does not depend on the line bundle $\mathcal{L}$ but just on
the requirement that $\gcd{(a_{i}, b_{i}(\mathcal{L}))}=1$.

For `honest' line bundles $\mathcal{K}$ (i.e.\ having isotropy data
$b_{i}(\mathcal{K})=0$ 
$\forall i$) the degree and first Chern
class agree
\be
c_{1}(\mathcal{K})=\deg{(\mathcal{K})} \nonumber
\ee
Moreover, if $\mathcal{L}$ is a $V$-line bundle and $\mathcal{K}$ a line
bundle, we have
\be
\deg{(\mathcal{K}\otimes \mathcal{L})} =
\deg{(\mathcal{K})}+\deg{(\mathcal{L})} \nonumber 
\ee
so that
\be
\deg{(\mathcal{L}^{\otimes n}\otimes \mathcal{K})} +
\deg{(\mathcal{L}^{-\otimes n}\otimes \mathcal{K}^{-1}) } =  - N +
\sum_{i=1}^{N}\phi_{a_{i}}(n) \label{sum-deg}
\ee
is independent of $\mathcal{K}$ and
\be
\deg{(\mathcal{L}\otimes \mathcal{K})}-\deg{(\mathcal{L}^{-1}\otimes
  \mathcal{K}^{-1})} = 2\, \deg{(\mathcal{K})} + 2\, c_{1}(\mathcal{L}) - 2
\sum_{i=1}^{N}((b_{i}(\mathcal{L})/a_{i})) \label{sub-deg}
\ee

\subsection*{The Principal Bundle Structure on $M$}

Let $\kappa$ be a connection on the principal $U(1)$ V-bundle
$\mathcal{L}$ that defines our 3-manifold
$M= S(\mathcal{L})$. We think of $\kappa$ as a globally defined real-valued 1-form on the 
total space of the bundle, 
and denote by $\xi$ the fundamental (or Reeb) vector field on $M$, i.e.\ the
generator of the $U(1)$-action. A $U(1)$ connection $\kappa$ is characterised
by the two conditions
\be
\iota_{\xi}\kappa = 1, \;\;\; L_{\xi} \, \kappa = 0
\label{equiv}
\ee
where $L_{\xi} = \{ d\, , \, \iota_{\xi} \}$ is the Lie derivative in the
$\xi$ direction which imply that $\iota_{\xi} d\kappa=0$, so
that the curvature 2-form $d\kappa$ of $\kappa$ is horizontal, as behoves
the curvature of a connection.

In local coordinates one has
\be
\kappa = d\theta + \beta\;\;,\label{kap}
\ee
where $\theta$ is a fibre coordinate, $0\leq \theta <1$, and
$\beta=\beta_{\mu}\, dx^{\mu}$ is a local representative on $\Sigma$ of the
connection $\kappa$ on $M$. 

Our orientation conventions \cite{BTcs} are such that
$d\kappa$ is minus the Euler class or first Chern class of the bundle
over $\Sigma$,
\be
c_{1}(\mathcal{L}) = \int_{\Sigma}-d\kappa \nonumber 
\ee
and since $M$ has $c_{1}(\mathcal{L})= (n + 
\sum_{i}b_{i}(\mathcal{L})/a_{i})$, we choose $\beta$ so that the
curvature 2-form satisfies 
\be
d\kappa = - (n + \sum_{i=1}^{N}\frac{b_{i}(\mathcal{L})}{a_{i}})\,
\pi^{*}(\omega)
\label{pomega}
\ee
for $\omega$ a unit normalised symplectic form on $\Sigma$.

For $c_{1}(\mathcal{L}) \neq 0$ a choice of $\kappa$ equips $M$ with 
a contact structure, such that $\kappa \wedge d\kappa$ is nowhere
vanishing on $M$. Indeed,
\be
\kappa \wedge d \kappa = -(n +
\sum_{i=1}^{N}\frac{b_{i}(\mathcal{L})}{a_{i}})\, d\theta \wedge  
\pi^{*}(\omega) 
\ee
is nowhere vanishing as required providing that the $U(1)$ V-bundle is
non-trivial. We also note that
\be
\int_{M} \kappa \wedge d \kappa =  - (n +
\sum_{i=1}^{N}\frac{b_{i}(\mathcal{L})}{a_{i}}) \int_{\Sigma} \omega =
-c_{1}(\mathcal{L}) \;\;.\label{p}
\ee

\section{Chern-Simons Theory on Seifert 3-Manifolds}
\label{seccs}

Much of the construction that we use has been explained in great
detail in \cite{BTcs} so we will be very brief about it here.
We fix the gauge group $G$ to be compact, semi-simple, and simply
connected so that the principal $G$-bundle on the
3-manifold $M$ and all its associated vector bundles are trivial. 
In principle the extension to trivial bundles for non-simply connected
groups is reasonably straightforward (and will mainly lead to a few extra signs
in the formulae), while the extension to non-trivial bundles of non-simply
connected groups requires some more thought in relation with diagonalisation 
and the argument based on the Gysin sequence.

Given that $\kappa$ is nowhere vanishing, any one-form $\beta \in
\Omega^{1}(M, \mathbb{R})$ may be decomposed as
$\beta = \beta_{\kappa} + \beta_{H}$ with 
\be
\beta_{\kappa}=\kappa \wedge \iota_{\xi} \, \beta \in \Omega^{1}_{\kappa}(M,
\mathbb{R}),
\;\;\; \beta_{H} = (1-\kappa
\wedge \iota_{\xi}) \, \beta \in \Omega^{1}_{H}(M,\mathbb{R}).
\ee
One may also decompose connections, thought of as elements of
$\Omega^{1}(M, \lg)$,  
\be
A = A_{\kappa} + A_{H}\equiv  \phi \, \kappa +
A_{H}. 
\label{Adecomp} 
\ee
and as $\phi \in \Omega^{0}(M, \lg)$ it is correctly thought of as a
section of the adjoint bundle $E=M \times \lg$. 

The level $k$ Chern-Simons action is
\bea
kS_{CS}[A] &= &\frac{k}{4\pi}\int_{M} \Tr \left( AdA + \frac{2}{3} A^{3} \right)
\nonumber \\
& = & \frac{k}{4\pi}\int_M \Tr\left(
A_{H}\wedge \kappa \wedge L_{\phi}
\, A_{H} + 2\phi \, \kappa \wedge d\, A_{H} +
\phi^{2}\, \kappa\wedge d \, \kappa \right)\;\;.
\label{sm0}
\eea
We have changed notation from \cite{BTcs} for the Lie derivative to
$L_{\xi}=\{ \iota_{\xi} ,\, d\}$
and for the covariant Lie derivative to $L_{\phi}= L_{\xi}+ [ \phi,\, $ from $\mathcal{L}_{\xi}$ and $\mathcal{L}_{\phi}$ in
order to avoid conflict with our notation for bundles. 

\subsection*{Gauge Conditions}

We impose the gauge condition
\be
L_{\xi} A_{\kappa}=0 \Leftrightarrow 
L_{\xi}\phi = \iota_{\xi} \, d \,\phi = 0\;\;.
\label{gf}
\ee
This gauge condition, $L_{\xi}\phi=0$, tells us
that $\phi$ is a $U(1)$-invariant section of $E$. Equivalently, it can 
therefore be regarded as a section of the (trivial) adjoint V-bundle $V$
over $\Sigma$. Having pushed down $\phi$ to $\Sigma$ in this manner, we can now
proceed to the diagonalisation of $\phi$ as in \cite{btg/g}. Thus
let $T$ be some maximal torus of $G$ and $\lt$ the corresponding Cartan
subalgebra, with  $\lg = \lt \oplus \lk$ and set
\be
\phi^{\lk} = 0. \label{gfc}
\ee
As shown in \cite{btg/g,btdiag}, and discussed previously, there is a price
to pay for diagonalising sections of $V$ (in the sense of conjugating them into 
maps taking values in the Cartan subalgebra $\lt$).

Up to this point we have not imposed any particular conditions on
$M$. However, in order to determine the obstructions in a simple way
we ask that $M=S(\mathcal{L}_{0})$ with $\gcd{(a_{i},a_{j})}=1$ (for
$i\neq j)$ and that $c_{1}(\mathcal{L}_{0})= (\prod_{i} a_{i})^{-1}$
or that $M=S(\mathcal{L}_{0}^{\otimes d})$. With the condition that
$M=S(\mathcal{L}_{0}^{\otimes d})$ we must sum over all $T$-bundles on $M$
that one gets by pullback of certain $T$ bundles from $\Sigma$. The
line bundles on $\Sigma$ are generated by $\mathcal{L}_{0}$ (by
Theorem 2.3 of \cite{FS}), and as we have already explained, the ones
of interest to us are powers of
$\mathcal{L}_{P}=\mathcal{L}_{0}^{\otimes P}$ (\ref{P}). The pull-backs of the
$\mathcal{L}_{P}$ from $\Sigma$ to $M$ are of 
finite order and  all torsion
bundles on $M$ arise in this way, so that it is precisely these
bundles that we should sum over in the path integral.

\subsection*{More Conditions on $\phi$}

Those $A_{H}^{\lt}$ which are $U(1)$ invariant, $L_{\xi} \, A_{H}^{\lt}=0$,
do not appear in the kinetic term $A_{H}\wedge  L_{\phi}
A_{H}$ and so they can
only appear in the mixed kinetic term $2\phi \, \kappa \wedge  d\,
A_{H}$. The path 
integral over such $A_{H}^{\lt}$ then imposes the condition $
\iota_{\xi}d (\kappa \, \phi) = 0$.
This delta function constraint on $\phi$ together with the gauge
condition (\ref{gf}) imply that $\phi$ is actually constant,
\be
d \phi =0.\label{phiconstant}
\ee
Now with $\phi$ constant we have, with $M=S(\mathcal{L}_{0}^{\otimes d})$,
\be
\int_{M} \kappa \wedge d\kappa \,\Tr \,  \phi^{2} =  -\frac{d}{P}
\,\Tr \, \phi^{2} 
\ee

\section{Reduction to an Abelian Theory on $\Sigma$}\label{reduction}

Having discussed the effect of integrating out the $U(1)$-invariant modes
of $A_{H}^\lt$, we now keep these and investigate what happens upon 
integrating out the other modes and fields, with the understanding that
$\phi$ will ultimately turn out to be constant. All these fields appear
quadratically in the action, and therefore will give rise to ratios of
determinants. The definition and regularisation of these determinants
for $\Sigma$ smooth were specified in
detail in Appendix B of \cite{BTcs} we will take that for granted but
augment that discussion here to take the orbifold points into
account. 

Given the choice of metric
\be
g_{M} = \pi^{*}g_{\Sigma} \oplus \kappa \otimes \kappa
\ee
the operator $*\,\kappa \wedge \,
L_{\phi}$ acts on the space 
of  horizontal $\lk$-valued 1-forms,
\be
*\, \kappa \wedge \,  L_{\phi}:\Omega^{1}_{H}(M, \lk)
\rightarrow \Omega^{1}_{H}(M, \lk) .
\ee
Hence integrating over the $\lk$-components of the ghosts ghosts $(c^{\lk},
\overline{c}^{\lk})$ and the connection $A_{H}^{\lk}$, one obtains the
following ratio of determinants:
\be
\frac{\Det{ \left(
      iL_{\phi}\right)_{\Omega^{0}(M, \lk)}
  }}{\sqrt{\Det{\left(*\,  \kappa \wedge \,
       i L_{\phi}\right)_{\Omega^{1}_{H}(M, 
      \lk)} } }}\;\;. \label{ratio}
\ee

Integration over the ghosts $(c^{\lt}, \overline{c}^{\lt})$ and those
$A^{\lt}_{H}$ modes which are 
not $U(1)$ invariant give the following ratio of determinants:
\be
\frac{\Det^\prime{ \left(
      iL_{\xi}\right)_{\Omega^{0}(M, \lt)}
  }}{\sqrt{\Det^\prime{\left(*\,  \kappa \wedge \,
        iL_{\xi}\right)_{\Omega^{1}_{H}(M, 
      \lt)} } }} \label{ratio2}
\ee
The notation $\Det^\prime$ indicates that the zero
mode of the operator is not included. 

To evaluate these ratios of determinants we expand all the fields in
their Fourier modes in the $\xi$ direction. In particular for the
connection we set $A_{H} = \sum_{n=- \infty}^{\infty} A_{n}$
where the eigenmodes satisfy $L_{\xi}\, A_{n}= - 2\pi i n\,  A_{n}$ and $\iota_{K}\, A_{n} = 0$
and likewise for the ghosts $c$ and $\overline{c}$. These eigenmodes can
equivalently be regarded as sections of line bundles
$\mathcal{L}^{\otimes n}$ (where
$\mathcal{L}$ defines $M$) over $\Sigma$ (which pull back to the
trivial line bundle on $M$). Hence we have that
\be
\Omega^{0}(M, \mathbb{C}) = \bigoplus_{n}\, \Omega^{0}(
\Sigma,\mathcal{L}^{\otimes n} ),\label{decomp}
\ee
As we have singled out the Cartan subalgebra, the bundles that we are
working with effectively `split' so we think of the charged Lie
algebra valued forms on $M$ as sections of the trivial bundle $M
\times \lk$. In order to make a Fourier decomposition of such sections
we understand each mode to be a section of a trivial bundle $V_{\lk}$ on
$\Sigma$ which pulls back to $M
\times \lk$. Consequently, on tensoring (\ref{decomp}) with the
trivial bundles $V_{\lk}$ below and  
$\pi^{*}(V_{\lk}) = M \times \lk$ above we have
\be
\Omega^{0}(M, \lk) = \bigoplus_{n}\, \Omega^{0}( \Sigma,
\mathcal{L}^{\otimes n}\otimes V_{\lk})\;\;.\label{f2} 
\ee
A similar discussion shows that each mode $n$ of a horizontal 1-form on
$M$ is one to one with a section on $\Sigma$, consequently one has
\be
\Omega^{1}_{H}(M, \lk) = \bigoplus_{n} \, 
\Omega^{1}( \Sigma, \mathcal{L}^{\otimes n}\otimes V_{\lk}). \label{f1}
\ee
Now, as explained in \cite{BTcs}, the ratio of determinants
(\ref{ratio}) and (\ref{ratio2}) need a definition (and
regularisation). We set
\be
\sqrt{\Det{Q}} = \sqrt{\left|\Det{Q}\right|}\, \exp{\frac{+i\pi}{2} \,
  \eta(Q)}
\ee
where $
\eta(Q) = \frac{1}{2} \sum_{\lambda \in \mathrm{spec}(Q)} \mathrm{sign}(\lambda)$
and the root is the positive
root for either of the operators that appear in (\ref{ratio}) and
(\ref{ratio2}). We 
regularise the absolute value and the phase (assuming that zero is not
an eigenvalue) by setting
\bea
\left|\Det{Q}\right|(s) &=& \exp{\sum_{\lambda\in \mathrm{spec}(Q)}
  e^{ s \Delta} \ln 
  |\lambda|} \label{absdet}
\\
\eta(Q, \, s) &=& \frac{1}{2} \sum_{\lambda\in \mathrm{spec}(Q)}
\frac{\mathrm{sign}(\lambda)}{|\lambda|^{s} }\exp{s \Delta} \label{etas}
\eea
for $\Delta$ an appropriate negative definite operator. As explained
in \cite{btg/g} an appropriate choice of $\Delta$ is the Laplacian of
the twisted Dolbeault operator on $\Sigma$.

In order to state the results that we borrow from \cite{btg/g, BTcs} we need
to introduce some notation. Each charged section contributes to the
determinant but its contribution depends on the charge, so we decompose
the charge space into roots
\be
V_{\lk} = \oplus_{\alpha} V_{\alpha}
\ee
The regularisation that we have chosen then leads us to considering
the index of the Dolbeault operator (how this comes about can be found
around (6.14) of \cite{btg/g}). Now the Riemann-Roch-Kawasaki index
theorem for a line V-bundle $\mathcal{L}$ on an orbifold \cite{Kawasaki} states that
\be
\mathrm{Index}(\overline{\partial}_{\mathcal{L}}) \equiv \chi(\Sigma,
\, \mathcal{L}) \equiv
\dim_{\mathbb{C}}{\mathrm{H}^{0}(\Sigma,\, \mathcal{L})}-
\dim_{\mathbb{C}}{\mathrm{H}^{1}(\Sigma,\, \mathcal{L})}=
\deg{\left( \mathcal{L}\right)} + 1-g
\ee
and one should note that it is the degree that enters and not the
first Chern class.

Returning to the determinants, we find that as far as the norm is concerned it
reduces to
\be
\sqrt{\prod_{\alpha}\prod_{n}(2\pi n + i \alpha(\phi))^{\chi(\Sigma,
  \mathcal{L}^{\otimes n}\otimes V_{\alpha})-\chi(\Sigma, K_{\Sigma} \otimes
  \mathcal{L}^{\otimes n}\otimes V_{\alpha} )}}
\ee
with $K_{\Sigma}$ the canonical bundle of $\Sigma$. By Serre
duality we have that $\chi(\Sigma, K_{\Sigma}\otimes \mathcal{L}^{\otimes
  n}\otimes V_{\alpha})=
-\chi(\Sigma, \mathcal{L}^{\otimes -n}\otimes V_{-\alpha}) $ so the
exponent in the previous expression is 
\bea
\chi(\Sigma, \mathcal{L}^{\otimes n}\otimes
V_{\alpha})+\chi(\Sigma, \mathcal{L}^{\otimes
  -n}\otimes V_{-\alpha}) &=& 2-2g +
\sum_{n}\left[\deg{(\mathcal{L}^{\otimes n})} + 
\deg{(\mathcal{L}^{-\otimes n})} \right]\nonumber \\
& & \label{pluschis}
\eea
where we have made use of (\ref{sum-deg}). By inspection of
(\ref{pluschis}) one sees that the absolute value of the determinants
is the same for $S(\mathcal{L})$ and $S(\mathcal{L}^{-1})$.

The eta invariant of the phase of the determinant is by (B.26) of
\cite{BTcs} 
\bea
\eta(L_{\phi}, \,s)&= & \eta_{(0,1)}(i L_{\phi})(s) +
\eta_{(1,0)}(-i L_{\phi})(s) \nonumber \\
& =& -\frac{1}{2}\sum_{n , \; \alpha } \left(
  \chi(\mathcal{L}^{\otimes n}\otimes
  V_{\alpha}) + 
  \chi(K \otimes  
\mathcal{L}^{\otimes n}\otimes V_{\alpha} ) \right)  \frac{\mathrm{sign}(2\pi
n+
i\alpha(\phi))}{ |2\pi n+i\alpha(\phi)|^{s}}\nonumber\\
&=& -\frac{1}{2}\sum_{n , \; \alpha } \left( \chi(\mathcal{L}^{\otimes n}\otimes
  V_{\alpha})-  \chi(\mathcal{L}^{\otimes -n}\otimes
  V_{-\alpha})\right) \frac{\mathrm{sign}(2\pi
n+
i\alpha(\phi))}{ |2\pi n+i\alpha(\phi)|^{s}}
\eea
the last line following by Serre duality. The subscripts on the
$\eta$'s in (B.26) of \cite{BTcs} are there to indicate whether we are
using the index of $\partial$ or of $\overline{\partial}$.

Without loss of generality we choose $\phi$ such that 
$0 <i \alpha(\phi) < 2\pi$ for the positive roots, so that
\bea
\eta(L_{\phi}, s) & =& -  \sum_{\alpha >0}
[\deg{(V_{\alpha})}-\deg{(V_{-\alpha})} ]\, |i
\alpha(\phi)|^{-s} \nonumber \\
& & - \sum_{n\geq 1}\sum_{\alpha >0} 
    [\deg{(\mathcal{L}^{\otimes n}\otimes
      V_{\alpha})}-\deg{(\mathcal{L}^{\otimes 
        -n}\otimes V_{-\alpha})} ]\, (2\pi n+i
    \alpha(\phi))^{-s}  \nonumber \\
& & - \sum_{n\geq 1}\sum_{\alpha >0}[\deg{(\mathcal{L}^{\otimes n}\otimes
  V_{-\alpha})}-\deg{(\mathcal{L}^{\otimes 
        -n}\otimes V_{\alpha})} ]\,
     (2\pi n- i \alpha(\phi))^{-s} \nonumber
\eea
By (\ref{sub-deg}) we can split the phase as
\be
\eta(L_{\phi},s) = \sigma(L_{\phi},V_{\lk}, s) +
\gamma(L_{\phi},\mathcal{L}, s) \nonumber
\ee
where
\bea
\sigma(L_{\phi},V_{\lk}, s) & = & - 2 \sum_{\alpha >0}\deg{(V_{\alpha})}|i
\alpha(\phi)|^{-s}-2 \sum_{\alpha >0}\deg{(V_{\alpha})}\sum_{n\geq
  1}(2\pi n+i 
    \alpha(\phi))^{-s} \nonumber\\
& & \;\;\; + 2 \sum_{\alpha >0}\deg{(V_{\alpha})}\sum_{n\geq
  1} (2\pi n- i \alpha(\phi))^{-s}\label{etaV}
\eea
and
\bea
\gamma(L_{\phi},\mathcal{L},s) &= & - \sum_{n\geq 1}\sum_{\alpha >0} 
    [\deg{(\mathcal{L}^{\otimes n})}-\deg{(\mathcal{L}^{\otimes 
        -n})} ]\,\left[ (2\pi n+i
    \alpha(\phi))^{-s} + (2\pi n-i
    \alpha(\phi))^{-s}\right]\nonumber\\
& & \label{etaphi}
\eea
Now $\sigma(L_{\phi},V_{\lk}, s) $ does not depend explicitly on
$\mathcal{L}$ so, in particular, we would find the same result had
we used any other line V-bundle. However,
$\gamma(L_{\phi},\mathcal{L},s)$ is quite a different object depending 
explicitly on the line V-bundle defining $M$ and in fact we have
\be
\gamma(L_{\phi},\mathcal{L}^{-1},s) = -
\gamma(L_{\phi},\mathcal{L},s) \label{etasign} 
\ee
This is as far as we can go in this generality.

\subsection{Absolute Value of the Determinant}
In order to determine the absolute value of the determinants we use
\be
\chi(\Sigma, \mathcal{L}^{\otimes
  n})+\chi(\Sigma, \mathcal{L}^{\otimes -n}) & = & \chi(\Sigma, \mathcal{L}_{0}^{\otimes dn})
+  \chi(\Sigma, \mathcal{L}_{0}^{-\otimes dn}) \nonumber\\
&=& 
2-2g - N + \sum_{i=1}^{N}\phi_{a_{i}}(dn) \nonumber
\eea
The last line follows from the index theorem
as all our assumptions about the bundles for which $\phi_{a_{i}}(dn)$
is defined hold as we will now see.

We demand that the line bundle $\mathcal{L}=\mathcal{L}_{0}^{\otimes
  d}$ which defines our 3-manifold $M = S(\mathcal{L})$ has isotropy
invariants such that $\gcd{(a_{i}, b_{i}(\mathcal{L})
  )}=1$. However, we also have that $b_{i}(\mathcal{L}_{0}^{\otimes
  d}) = d b_{i}(\mathcal{L}_{0})\, \mod a_{i}$ so that
neither $d$ nor $b_{i}(\mathcal{L}_{0})$ are divisible by
$a_{i}$ otherwise $\gcd{(a_{i}, b_{i}(\mathcal{L})
  )} \neq 1$. We can do a bit better. Let $b_{i}(\mathcal{L}_{0}^{\otimes
  d}) = d b_{i}(\mathcal{L}_{0}) + m a_{i}$ for some
$m$. Suppose that $y\neq \pm 1$ divides $a_{i}$ then $y$ cannot
divide $d b_{i}(\mathcal{L}_{0})$ as that would conflict with our
assumption that $\gcd{(a_{i}, b_{i}(\mathcal{L})
  )}=1$. In particular $y$ cannot divide either $d$ or
$b_{i}(\mathcal{L}_{0})$, so we have that $\gcd{(a_{i},
  b_{i}(\mathcal{L}_{0})   )}=1$ and  $\gcd{(a_{i},
  d)}=1$. Consequently, we do indeed have that
\bea
\deg{(\mathcal{L}^{\otimes n})} +
\deg{(\mathcal{L}^{\otimes -n})} &
= &
\deg{(\mathcal{L}_{0}^{\otimes dn})} +\deg{(\mathcal{L}_{0}^{-\otimes
      dn})} \nonumber \\
& =&   - N + \sum_{i=1}^{N}\phi_{a_{i}}(dn)
 \eea
Are there non zero values of $\phi_{a_{i}}(dn)$ and if so what
form do they take? Since $\phi_{a_{i}}(dn)$ is non-zero only if
$a_{i}| dn$ and we know that $a_{i}$ does not divide $d$ it must
divide $n$. Whence only those $n=ma_{i}$ for $m\in \mathbb{Z}$ yield
non-zero $\phi_{a_{i}}(dn)$.

Now, up to normalisation,
\be
\prod_{\alpha}\prod_{n} (2\pi n + i \alpha(\phi)) \simeq
T_{S^{1}}(\phi)
\ee
where
\bea
T_{S^{1}}(\phi) = \det{}_{\lk}{(1- \Ad{ \,
    \ex{\phi}})}& =& \prod_{\alpha>0}(1- \ex{\alpha( \phi)})(1
-\ex{-\alpha(\phi)}) \nonumber\\
& = & \prod_{\alpha>0} 4 \sin^{2}{\left( i\alpha(\phi)/2\right)}\label{ts1}  
\eea
is the Ray-Singer torsion of $S^{1}$ (with
respect to the flat connection $ i\phi d\theta$).

We still need to determine
\be
\prod_{\alpha}\prod_{n} (2\pi n + i
\alpha(\phi))^{\phi_{a_{i}}(dn)}
\ee
As argued above, the function $\phi_{a_{i}}(dn)$ vanishes except when
$n = m a_{i}$ for $m \in \mathbb{Z}$ so that
\bea
\prod_{\alpha}\prod_{n} (2\pi n + i
\alpha(\phi))^{\phi_{a_{i}}(dn )} &=&
\prod_{\alpha}(a_{i})^{\sum_{n} 1}\prod_{m} (2\pi m 
+i\alpha(\phi))/a_{i}) \nonumber \\
&=&
\prod_{\alpha}\prod_{m} (2\pi m 
+i
\alpha(\phi))/a_{i}) \nonumber\\
&\simeq& T_{S^{1}}((\phi/a_{i})
\eea
where we have used the fact that the zeta function regularisation of
$\sum_{n=-\infty}^{\infty}1$ is zero.

Putting the pieces together for the absolute value of the determinant
we find it is just what we called $\sqrt{T_{M}}$ at the end of Section
\ref{sec2}, namely
\be
\sqrt{T_{M}(\phi, \, a_{1}, \dots , a_{N})}= T_{S^{1}}(\phi)^{1-g-N/2}. \prod_{i=1}^{N}
T_{S^{1}}(\phi/a_{i})^{1/2} \label{absval}
\ee
Notice that this does not depend on the weights
$b_{i}(\mathcal{L}_{0}^{\otimes d})$ (and seemingly nor on $d$, but as
we will see $\phi$ depends on $d$). So, in particular, (\ref{absval})
does not depend on the orientation of $M$ which we explained, in a
slightly different way, just after (\ref{pluschis}).

\subsection{Phase of the Determinant}

Nicolaescu \cite{Nic} has done some of the work for us. In
particular the trick of passing from $\lfloor x \rfloor$ to $\ldb x
\rdb$ we took from him and this allows us to write certain terms as
Dedekind sums later on. We will need to make use of the Hurwitz zeta function
\be
\zeta(s,x)= \sum_{m \geq 0} \frac{1}{(m+x)^{s}},  \nonumber
\ee
which for negative integral $s$ is related to the Bernoulli
functions. In particular,
\be
\zeta(0,x) = \frac{1}{2}-x,\;\; \mathrm{and}\;\; \zeta(-1,x)=
-\frac{x^{2}}{2} + \frac{x}{2} - \frac{1}{12} \nonumber
\ee
whence
\bea
\sum_{n\geq 1} \frac{1}{(2\pi n + i \alpha(\phi))^{s}} &=& - \frac{1}{2}
- \frac{i}{2\pi} \alpha(\phi) + \mathcal{O}(s)\nonumber\\
\sum_{n\geq 1} \frac{n}{(2\pi n + i \alpha(\phi))^{s}} &=& - \frac{1}{12}
- \frac{1}{8\pi^{2}} \alpha(\phi)^{2} + \mathcal{O}(s) \nonumber
\eea
Clearly for (\ref{etaV}) as $s \longrightarrow 0$,
\be
\sigma(L_{\phi},V_{\lk},s) = -2\sum_{\alpha >0} \deg{(V_{\alpha})}\left( 1 +
\frac{1}{\pi} i \alpha(\phi) \right) + \mathcal{O}(s)
\ee

In order to determine the phase of (\ref{etaphi})  we make
use of the difference of degrees formula
\be
\deg{(\mathcal{L}^{\otimes n} )} -
\deg{(\mathcal{L}^{-\otimes n})} =
2n.c_{1}(\mathcal{L}) - 2 \sum_{i=1}^{N}
\ldb\frac{n b_{i}(\mathcal{L})}{a_{i}}\rdb 
 \nonumber
\ee

We have two types of terms to compute, those proportional to
$c_{1}(\mathcal{L})$ and those proportional to the symbol
$((x))$. We start with the ones proportional to
$c_{1}(\mathcal{L})$, namely, from (\ref{etaphi}),
\bea
& & -2 c_{1}(\mathcal{L})\sum_{\alpha >0} \left(
  \sum_{n\geq 1} \left[\frac{n}{(2\pi n + i
        \alpha(\phi))^{s}}  + \frac{n}{(2\pi n - i
        \alpha(\phi))^{s}}\right]\right) \nonumber\\
& & \; = c_{1}(\mathcal{L})\sum_{\alpha >0} \left(
  \frac{1}{3} +\frac{1}{2\pi^{2}}\alpha(\phi)^{2}  \right) + \mathcal{O}(s)
\eea

Now the terms in (\ref{etaphi}) proportional to
$((.))$ are, with $b_{i}=b_{i}(\mathcal{L})$,
\be
2\sum_{\alpha >0}\sum_{i=1}^{N}
  \sum_{n\geq 1} \sum_{\pm}\left[\ldb\frac{nb_{i}}{a_{i}}\rdb
.\frac{1}{(2\pi n \pm i 
        \alpha(\phi))^{s}} \right]\nonumber
\ee
One can allow the sum over $n$ to include $n=0$ since the two
contributions cancel (as $s$ goes to zero). Furthermore, the
periodicity $((x+1)) = ((x))$ allows us to write
\bea
&&\sum_{n\geq 0}\ldb\frac{n 
      b_{i}}{a_{i}}\rdb
.\frac{1}{(2\pi n \pm i  \alpha(\phi))^{s}} \nonumber\\
& &\;\; =
      \sum_{k=0}^{a_{i}-1}\ldb\frac{k
      b_{i}}{a_{i}}\rdb
      \frac{1}{(2\pi a_{i})^{s}}
    \zeta\left(s, \frac{k \pm i \alpha(\phi)/2\pi}{a_{i}} \right)
      \nonumber \\
& & \;\; =\sum_{k=0}^{a_{i}-1}\left(\!\left(\frac{k
      b_{i}}{a_{i}}\rdb 
\left(\frac{1}{2}-\frac{k}{a_{i}} \mp 
      \frac{i\alpha(\phi)/2\pi}{a_{i}}\right) + \mathcal{O}(s)\nonumber\\
&  &\;\; = \sum_{k=1}^{a_{i}-1}\ldb\frac{kb_{i}}{a_{i}}\rdb
[-\ldb\frac{k}{a_{i}}\rdb
\mp 
      \frac{i\alpha(\phi)/2\pi}{a_{i}}] + \mathcal{O}(s)\nonumber
\eea
as $1/2-k/a_{i}=-((k/a_{i}))$ for $0 <k <a_{i}$. We have that 
\be
\sum_{n\geq 0} \sum_{\pm}\ldb\frac{n b_{i}}{a_{i}}\rdb
.\frac{1}{(2\pi n \pm i 
        \alpha(\phi))^{s}}=
      -2\sum_{k=1}^{a_{i}-1}\ldb\frac{k b_{i}}{a_{i}}\rdb
    .\ldb\frac{k}{a_{i}}\rdb  + \mathcal{O}(s) \nonumber
\ee

The Dedekind sum $s(b,a)$ is defined by
\be
s(b, a) = \sum_{k=1}^{a-1}\ldb
\frac{kb}{a} \rdb \ldb \frac{k}{a} \rdb \nonumber
\ee
whence
\be
\sum_{n\geq 0} \sum_{\pm}\ldb\frac{nb_{i}}{a_{i}}\rdb
.\frac{1}{(2\pi n \pm i 
        \alpha(\phi))^{s}}= -2s(b_{i}, a_{i})+ \mathcal{O}(s)\nonumber
\ee

Putting the pieces together, thus far we have
\bea
\eta(L_{\phi},0) & =& \sum_{\alpha >0} \left(
  -2c_{1}(V_{\alpha})  +
  \frac{c_{1}(\mathcal{L})}{3} +
  \frac{c_{1}(\mathcal{L})}{2\pi^{2}}\alpha(\phi)^{2} +
  \frac{2c_{1}(V_{\alpha})}{\pi} 
  i\alpha(\phi) \right)\nonumber \\
& & \;\;\;\;\;  - 4 \sum_{i=1}^{N}\sum_{\alpha >0}s(b_{i}, a_{i})
\eea

We also have to consider the contribution from the fields lying in the
Cartan subalgebra and these couple neither to the bundles $V_{\alpha}$
nor to the $\phi$. They contribute,
\be
\eta(L_{\xi},0) = \dim{T}\left( \frac{c_{1}(\mathcal{L})}{6} - 2 s(b_{i},
  a_{i}) \right)
\ee
Collecting all the contributions we find that the total $\eta(s)=
\eta(L_{\phi},s) + \eta(L_{\xi},s)$, as $s \longrightarrow 0$ is 
\bea
\eta(0) & = &   \sum_{\alpha >0} \left( -2c_{1}(V_{\alpha}) +
  \frac{c_{1}(\mathcal{L})}{2\pi^{2}}\alpha(\phi)^{2} +
  \frac{2c_{1}(V_{\alpha})}{\pi} 
  i\alpha(\phi) \right)\nonumber \\
&& \;\;\; + \dim{G}\left(   \frac{c_{1}(\mathcal{L}) }{6} - 
  \sum_{i=1}^{N}s(b_{i},  a_{i}) \right) \label{etalast} 
\eea
so that in this case we have
\be
-\frac{i\pi}{2} \eta(0) = 4\pi i \Phi(\mathcal{L}) -  \frac{ic_{\lg}}{4\pi} \int_{\Sigma} \left(
  \frac{d}{P}\Tr \phi^{2} \, \omega\right)+ \frac{ic_{\lg}}{2\pi}
\int_{\Sigma} \Tr \phi F_{A} \label{total-phase}
\ee
where
\be
\Phi(\mathcal{L}) = \frac{1}{48} \dim{G}\,\left(
  12\sum_{i=1}^{N}  s(b_{i},
  a_{i}) - \frac{d}{P}\right) \label{phase-phi}
\ee
As stated at the beginning of section \ref{seccs}, we consider the case
that $G$ is simply-connected. 
For such groups one has $\sum_{\alpha >0} c_{1}(V_{\alpha})
\in 2 \mathbb{Z}$ (the Weyl vector is integral), so that this term 
does not contribute to the phase.

It is important to notice that had we used the line V-bundle
$\mathcal{L}^{-1}$ rather than $\mathcal{L}$ to construct $M$
then the Chern class of $M$ would change sign
$c_{1}(\mathcal{L}) \rightarrow 
c_{1}(\mathcal{L}^{-1}) = - c_{1}(\mathcal{L})$ and so too the
Dedekind sum $s(b_{i},
a_{i}) \rightarrow s(a_{i}-b_{i},  a_{i}) = -s(b_{i},  a_{i}) $ as
required by (\ref{etasign}).

The expression (\ref{total-phase}) agrees with that obtained in \cite{BTcs} for the Lens
spaces $L(p,1)$ on setting $dc_{1}(\mathcal{L}_{0})=-p$, which can be
achieved by taking the $a_{i}=1$ and $b_{i}=0$ for all $i$ (so that $P=1$),
$\deg{\mathcal{L}_{0}}=1$ and then $d=-p$.

\section{Evaluating the Path Integral on $\Sigma$}

Now that we have integrated over all the $\lk$-valued fields as well as all the
$\lt$-valued modes which are not constant along the $S^{1}$ fibres in $M$,
the Chern-Simons partition function, up to the phase
(\ref{phase-phi}),  reduces to a 
path integral of an Abelian 2-dimensional gauge theory on
$\Sigma$ with action 
\be
S_M\ra S_{\Sigma}[A_H,\phi]= \frac{k+ c_{\lg}}{4\pi} \int_{\Sigma} \Tr \, 
(2\phi \, F_{H} -\frac{d}{P} \phi^{2} \omega)\;\;,
\label{ssigma}
\ee
where $A_H=A_H^{\lt}$ and $\phi=\phi^{\lt}$. 

The curvature 2-form $F_{H}$ includes the contribution of non-trivial
line bundles (but not line V-bundles) on $\Sigma$. To incorporate
those we let
\be
F_{H} \rightarrow F_{H}(A) + 2\pi \, r \, \omega \nonumber
\ee
with $r \in \mathbb{Z}^{\mathbf{rk}}$. $A_{H}$ is now understood to
be a $\lt$ valued 1-form on $\Sigma$. It might appear to be more
natural to have chosen that $r \in \mathbb{Z}_{d}$ as we did in
\cite{BTcs}.  However, as we have
seen previously \cite{BTcs} and as is also evident from the various
equalities in (\ref{z}), the net effect is the same, so 
for the sake of variety here we choose 
$r \in \mathbb{Z}^{\mathbf{rk}}$.

Baily \cite{Baily} tells us that
on an orbifold $\Sigma$ the Hodge decomposition is still available
where all sections are understood in the appropriate sense. Locally around an orbifold point (the conic point of $D/\mathbb{Z}_{a}$)
any p-form $\alpha$ is understood to be a $\mathbb{Z}_{a}$ invariant
p-form on $D$. Then, for two such 1-forms 
\be
\int_{D^{2}/\mathbb{Z}_{a}} \alpha \wedge \beta =
\frac{1}{|a|}\int_{D^{2}} \alpha \wedge \beta \nonumber
\ee
and so on.

The Hodge decomposition tells us that the harmonic modes of
$A_{H}^{\lt}$ only contribute to the normalisation of the path
integral. The exact components are in the gauge directions of the
residual $U(1)^{\mathbf{rk}}$ gauge symmetry and so may be set to zero
by a gauge choice. The co-exact parts of
$A_{H}^{\lt}$ are the only pieces that appear in the action and
integrating over those imposes the condition 
\be
d\phi^{i} =0\;\;.\nonumber
\ee
so that the $\phi$ are constant. With $\phi$ constant, the partition
function reduces to the finite-dimensional integral over the Cartan subalgebra
\bea
& & Z_k[M, G] \sim
\ex{4\pi i \Phi(\mathcal{L})}
\sum_{r \in \mathbb{Z}^{
    \mathbf{rk}}} \,  \int_{\lt} 
\sqrt{T_{M}(\phi)}\, \exp{\left(i \frac{k+
      c_{\lg}}{4\pi} \Tr \left( -\frac{d}{P} \, \phi^{2} + 4 \pi r\, \phi
  \right)\right)} \nonumber\\
& & \label{final1}
\eea

However, this is not the final form of the partition function as there
is still a discrete symmetry that we should mod out by. The partition
function (\ref{final1}) has the form of the gauge invariant partition
function (\ref{parent}) and so is invariant under the action of the affine Weyl
group $\Gamma^{W}$. How does this symmetry arise in the present
situation? 
\begin{itemize}
\item
Invariance under the integral lattice $I$ is there since, as we
have already noted, the pullback of any multiple of
$\mathcal{L}_{0}^{P}$ to $M$ is trivial so all such multiples are
equivalent. Explicitly we made the substitution
\be
A = A_{H} + \phi \, \kappa + 2\pi r \frac{P}{d} \, \kappa\nonumber
\ee
which obviously has the symmetry
\be
\phi \rightarrow \phi - 2\pi P s, \;\;\; r \rightarrow
r+ ds \nonumber
\ee
\item
The Weyl group makes an appearance since it was part of
the original $G$ symmetry. It acts, therefore, by
conjugation and on $\lt$ this becomes permutation of the (diagonal)
matrix entries. Permutation of both the $\phi$
and $r$ entries in the same way leaves $\Tr{(\phi^{2})}$ and
$\Tr{(r.\phi)}$ invariant. The Ray-Singer torsion of the circle is
also invariant under $W$ so we have that theory posses the symmetry
that we claimed. 
\end{itemize}
The partition function is, therefore,
\bea
Z_k[M, G] =  \Lambda\,\ex{4\pi i \Phi(\mathcal{L})}\,  \sum_{r \in
  \mathbb{Z}^{ 
    \mathbf{rk}}} \,  \int_{\lt/\Gamma^{W}} 
\sqrt{T_{M}(\phi)}\, \exp{\left(i \frac{k+
      c_{\lg}}{4\pi} \Tr \left( -\frac{d}{P} \phi^{2} + 4 \pi r\, \phi
  \right) \right)}  && \nonumber \\
&&\label{xfinal2a}
\eea
where $\Gamma^{W} = I \rtimes W$ is the affine Weyl group and $\Lambda$ is
a real normalisation constant that remains to be determined.

As the Ray-Singer torsion has zeros at the boundary of the Weyl
chamber the integrals (\ref{final1},
\ref{xfinal2a}) diverge when $g+N/2 >1$. As shown in \cite{btg/g} for
the smooth case one ought to regularise by giving a small mass term to
the connection, while preserving the residual $U(1)^{\mathbf{rk}}$
invariance. The same regularisation is applicable when $\Sigma$ is an
orbifold and guarantees the vanishing of the ghost determinant at the
boundary while the inverse of the determinant coming from the
connection remains finite. The net effect of this procedure is to exclude the
boundaries of the Weyl chamber.  As the contributions to the path
integral are at 
discrete points this regularisation prescription renders the integrals finite.

Witten \cite{WCS} shows how at one loop level the Chern-Simons
partition function becomes an integral over the moduli space of flat
connections with measure the square root of the Ray-Singer
Torsion. There is also a
phase factor coming from the Chern-Simons function and a framing
correction. We note that (\ref{xfinal2a}) has precisely the form just
described with $T_{M}(\phi,;\, a_{1} \dots , a_{N})$ being the
Ray-Singer Torsion of a Seifert $\mathbb{Q}$[g]HS but with a crucial
difference. Rather than the moduli space of flat connections on $M$ we have
instead the integral over $\lt/ \Gamma^{W}$ coming from the vertical part
of the connection. This is a much simpler integral to perform.

\section{The Inclusion of Wilson Loops along the Fibre}

We can also easily evaluate Wilson lines which are in the fibre
direction of $M$ thought of as a principal bundle. Such Wilson lines
only depend on the representation and on the $\kappa \phi$ part of the
connection. Since they do not depend on $A_{H}$ the inclusion of
Wilson loops does not change any of the arguments in the evaluation of
the path integral. In particular one may just as well take $\phi$ to
be constant and to take values in $\lt$.

The expectation value (normalised or not) of such a Wilson
loop $
\Tr_{R_{j}}{\left(\mathrm{P}\exp{\left(\oint
        A\right)}\right)}$
then is the same as evaluating (\ref{xfinal2a}) with
\be
\Tr_{R_{j}}{\left(\exp{\left(\oint \kappa
        \phi\right)}\right)} \label{Wilson-loop} 
\ee
inserted in the integral (or products of these). Now providing the
fibre is not exceptional (i.e.\ it is based at a 
regular point on the orbifold) the Wilson loop (\ref{Wilson-loop}) is $
\Tr_{R_{j}}{\left(\exp{\left(
        \phi\right)}\right)}$. The path integral including such Wilson
lines becomes,
\be
\Lambda\,\ex{4\pi i \Phi(\mathcal{L})}\,  \sum_{r \in
  \mathbb{Z}^{ 
    \mathbf{rk}}} \,  \int_{\lt/\Gamma^{W}} 
\sqrt{T_{M}(\phi)}\, \exp{\left(i \frac{k+
      c_{\lg}}{4\pi} \Tr \left( -\frac{d}{P} \phi^{2} + 4 \pi r\, \phi
  \right) \right)} \prod_{i} \Tr_{R_{j}}{\left(\exp{\left(
        \phi\right)}\right)}\nonumber
\ee
Had any of the Wilson loops been along an exceptional fibre (one which
is based at an 
orbifold point of weight $a_{i}$ on $\Sigma$), in the Wilson loops 
$\phi$ would have to be replaced by $\phi \ra \phi/a_i$.

\subsubsection*{Acknowledgement}
George Thompson would like to thank Chris Beasely for pointing out the paper
by Nicolaescu \cite{Nic} during the 2009 Chern-Simons Gauge Theory
meeting in Bonn.

\appendix

\section*{Appendices}
\section{An Example of Diagonalisation}

In this appendix we wish to explain in some more detail the appearance
of non-trivial T-bundles on diagonalisation. This is both a summary
and an extension to the orbifold case of one of the arguments given in
\cite{btdiag} for this phenomenon.

Let $M=S(\mathcal{L})$ be the circle bundle of a V-line bundle over an
orbifold $\Sigma$ and consider the (trivialised) vector bundle
$\ad{G}= M \times \lg$ over $M$. We equip $M$ with an induced contact
structure and let $\xi$ be the Reeb vector field on $M$. Now with
$\phi$ a section of $\ad{G}$ we mean $s : M  \longrightarrow M
\times \lg$ such that on $M$ it is the identity map. In this case a
section is the combination $s \cong (\mathrm{Id}_{M}, \, \phi)$ where
$\phi:M \longrightarrow \lg$. Now restrict
attention to those sections which satisfy $\iota_{\xi}d \phi =0$. Such
$\phi$ are still just maps to $\lg$. Or put another way, $s$ is still
a section of the trivial bundle $\ad{G}$ on $M$. 

However, such a
$\phi$ also defines a section $\hat{s}: \Sigma \longrightarrow \Sigma
\times \lg$ with the map automorphism on $\Sigma$ being the
identity map $\mathrm{Id}_{\Sigma}$. We are still dealing with a
trivial bundle albeit over an orbifold. In this context $\phi$ is a
map from $\Sigma$ to $\lg$. Notice, that at this point, the
information about the V-line bundle 
$\mathcal{L}$ (i.e.\ its isotropy weights $b_{i}$) no longer appears
and so too then information about $M$ is lost. (This is just as it is
in the case of Lens spaces $L(p,1)$ as for all of them the base is $S^{2}$.)

In case that $\Sigma = \widehat{\Sigma}/ \Gamma$ the map $\phi$ is equivalent to a
$\Gamma$ invariant 
map from $\widehat{\Sigma}$ to $\lg$. A good example of this situation
is the orbifold $S^{2}/\mathbb{Z}_{a}$ which has two marked points
both with isotropy $a$. Now a section of the trivial $\lg$ bundle over
$S^{2}/\mathbb{Z}_{a}$ is the same as an $\mathbb{Z}_{a}$ invariant
section of the trivial $\lg$ bundle over $S^{2}$. If, for $\zeta \in
\mathbb{Z}_{a}$, the section were simply equivariant $\phi(\zeta.z)=
\zeta.\phi(z)$  where the action on the Lie algebra is non-trivial,
then one would have a non-trivial $\lg$ bundle over $S^{2}/\mathbb{Z}_{a}$.

Let $\phi$ then be a
map from $S^{2} \rightarrow \lg$. For simplicity we take $\lg$ to be
$su(2)$. Given a connection on the bundle we can define an invariant
\be
n(\phi, \, A) = \frac{1}{2\pi i}\int_{S^{2}} \Tr{\left( \phi F_{A}
    -\frac{1}{4} \phi\, d_{A} \phi \wedge d_{A}\phi \right) }\label{n}
\ee
for those $\phi$ such that $\phi^{2} = -\mathrm{Id}_{2\times 2}$
(these are maps to $S^{2}$). This
invariant exhibits the non-trivial line bundles that arise on
diagonalisation. On the one hand the maps of interest  are maps
from $S^{2}$ to $S^{2}$ and so they fall into homotopy $\pi_{2}$
classes. In that case $n(\phi,0)$ just measures the winding number of
the map. Upon diagonalisation, with group map $g$, the map $g^{-1}\phi
g$ is
just to a single point 
and $n(g^{-1}\phi g, 0 + g^{-1}dg)$ is the first Chern class of the connection
$A=0+g^{-1}dg $. As $n(\phi, \, A)$ is gauge invariant the winding
number of $\phi$ and the first Chern class of the liberated line
bundles must agree. 

We are interested in maps to $S^{2}$ which are $\mathbb{Z}_{a}$
invariant. Embed $S^{2}$ in $\mathbb{R}^{3}$ so
that it is the solution to $x^{2}+y^{2}+z^{2}=1$. The action of
$\mathbb{Z}_{a}$ is
$\zeta.(w,z)= (\zeta.w,z)$ where $w=x+iy=
\exp{(i\theta)}. \sin{\varphi}$, $z=\cos{\varphi}$, where
$0\leq \theta < 2\pi$, $-\pi \leq \varphi \leq 0 $ and $\zeta =\exp{(2\pi
  i  /a)}$. As part of our example let
\be
\phi(\theta,\varphi) = \sin{\theta}.\sin{\varphi} . i\sigma_{1} +
\cos{\theta}.\sin{\varphi}.i\sigma_{2} + \cos{\varphi}.i\sigma_{3} 
\ee
be the identity map. The identity map is not $\mathbb{Z}_{a}$
invariant since the
action of $\mathbb{Z}_{a}$ is 
\be
\zeta:(\theta, \varphi) \longrightarrow (\theta + \frac{2\pi}{a},
\varphi)
\ee
However, it is quite straightforward to create maps which are
$\mathbb{Z}_{a}$ invariant, these are
\be
\widetilde{\phi}(\theta,\varphi) = \sin{(an\theta)}.\sin{\varphi} . i\sigma_{1} +
\cos{(an\theta)}.\sin{\varphi}.i\sigma_{2} +
\cos{\varphi}.i\sigma_{3} \label{winding-maps} 
\ee

It will not come as a surprise that their winding numbers
are elements of $a\mathbb{Z}$, indeed
\be
\frac{-1}{8\pi i}\int_{S^{2}} \Tr{\left(\widetilde{\phi}\, d
      \widetilde{\phi}
    \wedge d\widetilde{\phi}  \right)} = an
\ee
On $S^{2}/\mathbb{Z}_{a}$ the maps $\widetilde{\phi}$ become `winding'
number $n$ maps to the 2-sphere. On diagonalising (\ref{n}) tells us
that the first Chern class of the liberated line bundle is integral
and it is over such line bundles that we must sum.

\section{The Fundamental Group and Representations in $SU(2)$}
Our evaluation of the path integral does not
involve the moduli space of non-Abelian flat connections as
diagonalisation forces us to consider Abelian connections. 
By way of example we exhibit a non-trivial irreducible $SU(2)$ flat connection
on the Poincar\'{e} $\mathbb{Z}$HS in order to convince the reader
that such connections are there even though we manage to sidestep
having to face their existence in the course of our evaulation of the 
path integral.

To do this we start with the presentation of the generators and
relations of the fundamental group $\pi_{1}(M)$ for $M$ a $\mathbb{Q}$HS. After
the example we move on to determine the first cohomology group of these
manifolds as this is what really enters in the body of the paper. We
use the fact that the first homology group is the abelianisation of
the fundamental group $\mathrm{H}_{1}= \pi_{1}/[\pi_{1},\pi_{1}]$.

As we are
in the situation where $g=0$ the generators of $\pi_{1}(M)$ are $c_{j}$,
$j=1, \dots , 
N$ and $h$ subject to the relations $[c_{j}, h] =1$ and
\be
 c_{j}^{a_{j}}h^{b_{j}}=1, \;\;\; \prod_{j=1}^{N}
c_{j} = h^{n} \label{fund-rels}
\ee
where, $n$ is related to the degree of the line V-bundle that defines
$M$ and, $h$
is central. If $h=1$ the relations are just those for the fundamental
group of the orbifold $\Sigma$ so that $h$ is the generator along
the fibre. 

We give an example of a representation of $\pi_{1}$ for the
Poincar\'{e} 3-sphere $\Sigma(2,3,5)$ in $SU(2)$. The presentation
of $\pi_{1}$ is given by
\be
X^{2}= H^{-1}, \;\; Y^{3}= H^{-1} , \;\; Z^{5}= H^{-1}, \;\; XYZ =
H^{-1} \label{pi-in-su2}
\ee
and $H$ commutes with $X$, $Y$ and $Z$. We take $H$ to be
central (indeed one can show that for an irreducible representation it
must be) and for concreteness let $H=-\mathbb{I}_{2}$. We diagonalise
$Z$
\be
Z = \exp{\left( im\pi/5\, . \, \sigma_{3} \right)}, \;\;\; m =1,3, 5
\ee
(one is quickly led to a contradiction if one takes $Z$ central).
If we write
\be
X= a\mathbb{I}_{2}+ i \underline{b}.\sigma ,\;\;\; a^{2} +
|\underline{b}|^{2}=1
\ee
then the condition on $X$ in (\ref{pi-in-su2}) implies $a=0$, so $X
\in S^{2}$. We may
still act by conjugation by elements in the torus defined by
$\sigma_{3}$ without changing $Z$ and so we may rotate $X$ into an
$S^{1}$ of our choice, i.e.\ we simply set $b_{1}=0$, $b_{2}>0$ and we have that
\be
X= ib_{2}\sigma_{2} + ib_{3}\sigma_{3} 
\ee
We write $Y$ in the same way as we did $X$
\be
Y = c\mathbb{I}_{2}+ i \underline{d}.\sigma ,\;\;\; c^{2} +
|\underline{d}|^{2}=1
\ee
then the fact that $Y^{3}=-\mathbb{I}_{2}$ implies that $c=1/2$. The
only relation still to satisfy is $XYZ = -\mathbb{I}_{2}$. This is
straightforward and we find, with $\lambda_{0}= \cos{(\pi m/5)}$ and $\lambda_{1}=-\sin{(\pi m/5)}$,
\be
b_{3} = \frac{1}{2\lambda_{1}} \;\; d_{3} = -
\frac{\lambda_{0}}{2\lambda_{1}}  , \;\; d_{1}= b_{2}\lambda_{1}, \;\;
d_{2} = -b_{2}\lambda_{0} 
\ee
We now have a point in the space of flat $SU(2)$ connections on
$\Sigma(2,3,5)$. The matrices $X$, $Y$ and $Z$ are essentially the
holonomies of the flat connection in
question around the non-trivial cycles of $\Sigma(2,3,5)$.

As in \cite{BW} let $c_{N+1}=h$ so that the relations in $\mathrm{H}_{1}(M,
\mathbb{Z})$ may be written as
\be
\prod_{j=1}^{N+1} c_{j}^{A_{jk}} = 1 \label{constraint}
\ee
where
\be
A = \left(
\begin{array}{cccc}
a_{1} & 0 & \cdots  & b_{1}\\
0 & \ddots &0 & \vdots \\
0 & \cdots &  a_{N} & b_{N}\\
1  & \cdots & 1 & -n
\end{array}
\right) \nonumber
\ee
If $v \in \mathrm{H}_{1}(M,
\mathbb{Z})$ then $v^{d}$ must be trivial (in the multiplicative sense
so that $v^{d}=1$). An element $w=\prod_{j}c_{j}^{m_{j}}$ is trivial
iff $m_{j} = A_{jk} l_{k}$ and will be of the form $v^{d}$ providing
$\Det{A} \propto d$.

Calculating, we find that
\be
\Det{A} = -P\left( n + \sum_{i=1}^{N} \frac{b_{i}}{a_{i}} \right) 
\ee
We have then that $|\Det{A}|= |d|$ is the order of
$\mathrm{H}_{1}(M)$. One can be rather more explicit about this. Going back to
(\ref{fund-rels}) we have
\be
c_{j}^{P}h^{Pb_{j}/a_{j}}=1, \;\;\; \prod_{j=1}^{N}
c_{j}^{P} = h^{nP} \label{fund-rels2}
\ee
plugging the first into the second gives $h^{d}=1$. As we have
abelianised $\pi_{1}(M)$ to pass to $\mathrm{H}_{1}(M)=
\pi_{1}(M)/[\pi_{1}(M), \pi_{1}(M)]$ we may as well represent the generators as
\be
c_{j} = \exp{\left(2\pi i\, \frac{b_{j}}{a_{j}}.\frac{P}{d} \right)}, \;\;\; h=
\exp{\left(-2\pi i\, \frac{P}{d} \right)} \nonumber
\ee

\rnc{\Large}{\normalsize}

\end{document}